\documentclass[3p]{elsarticle}

\usepackage{amssymb,amsmath}
\usepackage{bm}
\usepackage{epsfig}
\usepackage{verbatim}
\usepackage{graphicx}
\usepackage{subcaption}

\usepackage{grffile}	
\usepackage{xcolor}
\usepackage[colorlinks]{hyperref}
\usepackage[colorinlistoftodos]{todonotes}

\usepackage{lineno,hyperref}
\modulolinenumbers[5]

\journal{Physica A: Statistical Mechanics and its Applications}

\newcommand{\beq}{\begin{equation}}
\newcommand{\eeq}{\end{equation}}
\newcommand{\beqa}{\begin{eqnarray}}
\newcommand{\eeqa}{\end{eqnarray}}
\newcommand{\beqann}{\begin{eqnarray*}}
\newcommand{\eeqann}{\end{eqnarray*}}

\begin{document}

\title{Pair correlations and structure factor of the $J_1$-$J_2$ square lattice Ising model in an external field}

\author[1]{Alejandra I. Guerrero \corref{cor1} \fnref{fn1}}
\ead{alejandra.i.guerrero@gmail.com}
\author[3]{Daniel A. Stariolo \fnref{fn2}}
\ead{stariolo@if.uff.br}

\cortext[cor1]{Corresponding author}
\fntext[fn1]{Present Address: Facultad de Ingenier\'{\i}as, Corporaci\'{o}n Universitaria Aut\'{o}noma del Cauca, CP 190003, Popay\'an, Colombia.}
\fntext[fn2]{Present Address: Departamento de F\'{\i}sica,
Universidade Federal Fluminense, 24210-346 Niter\'oi, Brazil.}

\address[1]{Departamento de F\'{\i}sica, Universidade Federal do Rio Grande do Sul, CP 15051, 91501-970, Porto Alegre, RS, Brazil}
\address[3]{Departamento de F\'{\i}sica,
Universidade Federal do Rio Grande do Sul and
National Institute of Science and Technology for Complex Systems, CP 15051, 91501-970 Porto Alegre, RS, Brazil}

\date{\today}

\begin{abstract}
We compute the structure factor of the $J_1$-$J_2$ Ising model in an external field on the square lattice
within the Cluster Variation Method. We use a four point plaquette approximation, which is the minimal one
able to capture phases with broken orientational order in real space, like the recently reported Ising-nematic
phase in the model. The analysis of different local
maxima in the structure factor allows us to track the different phases and phase transitions against temperature
and external field. Although the nematic susceptibility is not directly related to the structure factor,
we show that because of the close relationship between the nematic order parameter and the structure factor, the
latter shows
unambiguous signatures of the presence of a nematic phase, in agreement with results from direct minimization
of a variational free energy. The disorder variety of the model is identified and the possibility that the CVM
four point approximation be exact on the disorder variety is discussed.
\end{abstract}


\maketitle

\section{Introduction}
\label{Intro}
The structure factor, being a quantity of direct experimental access by neutron scattering and many other
spectroscopic techniques, is a central quantity in condensed 
matter physics~\cite{Ashcroft,ChLu1995}. Mathematically, it is the Fourier transform of the connected pair
correlation function, and as such its knowledge gives direct access to fluctuations and phase transitions
associated to them. Computing the structure factor then amounts to compute correlation functions, which is
known to be a hard task in statistical mechanics models. In order to characterize a phase transition it is
often possible to look at simpler one-point quantities, typically order parameters, like the magnetization
or the density. A qualitative understanding of a phase transition can be obtained by simple
mean field approximations. If one wants to compute universal quantities, like critical exponents, then it
is necessary to go beyond mean field approximations, for example through a Renormalization Group analysis.
But there are special kinds of order which are essentially associated with fluctuations and then, even if
one is interested in a qualitative description, simple mean field theory does not work. This is the case,
e.g. of broken orientational phases in systems with competing interactions~\cite{SeAn1995,StBa2010}. When
a competing attraction and repulsion or ferromagnetic and anti-ferromagnetic interactions are simultaneously
present, the system can develop modulated structures in the form of stripes or bubbles~\cite{SeAn1995}. 
These structures break rotational symmetry of space but may not break translational symmetry,
giving rise to phases with intermediate (in temperature or external field), purely orientational or nematic-like order, in analogy with the
nematic phases of liquid crystals~\cite{deGPr1998,ChLu1995}. Anisotropic phases with nematic-like order are
relevant, e.g. in low dimensional systems like electronic liquid-crystals~\cite{KiFrEm1998,BeFrKiTr2009,Fernandes2012,BaSt2009} 
and ultrathin ferromagnetic films~\cite{AbKaPoSa1995,CaMiStTa2006,SaRaViPe2010,NiSt2007}. 

In order to
characterize these nematic-like phases from microscopic models it is necessary to go beyond naive mean field
approximations. In particular, the orientational or nematic order parameter in two dimensional modulated systems is proportional
to the difference between correlation functions in two orthogonal space directions~\cite{StBa2010,BaSt2011}. As will be better described
below, a nematic order parameter can be defined as a weighted integral of the structure factor. Then, 
computing the structure factor gives direct access to the nematic-like order in systems with competing interactions
and orientational order. 

In a lattice, a systematic way of obtaining better approximations for the thermodynamics of a system is to consider
clusters of increasing size exactly. The Cluster Variation Method (CVM) is one of a family of cluster techniques
~\cite{Kikuchi51,Morita72,An88,Tanaka2002}.
Although of mean field character, it allows to improve considerably the locus of phase transition lines, specially
for systems with competing interactions where naive mean field usually gives a very poor approximation to the phase
diagram. It is also
suitable for computing approximations to multipoint correlation functions in a systematic way. The CVM has been
applied previously to compute the structure factor of a few models as the ferromagnetic Ising model~\cite{Sanchez82} and
the two dimensional ANNNI model~\cite{FineldeFontaine86}. 
In reference \cite{Cirillo99} the authors introduced a general approach for the computation of the structure
factor within the Cluster Variation Method, and applied it to the Ising model with nearest neighbors (NN), next-nearest 
neighbors (NNN) and plaquette interactions in two and three dimensions. For the case of NN and NNN interactions in the
square lattice, the so called $J_1$-$J_2$ Ising model, they computed the structure factor at zero external field in the paramagnetic phase. 
The phase transition lines between paramagnetic, ferromagnetic and collinear (stripe) phases where characterized and the
presence of a disorder line in the paramagnetic phase was obtained within the approximation and discussed in relation to the
exactly known result~\cite{Pelizzola2000}. Interestingly, in \cite{Pelizzola2000}, the four point CVM approximation was
proved to render the exact solution of the model at zero external field. In a recent work, we applied the CVM
to the $J_1$-$J_2$ Ising model in an external field~\cite{Guerrero15} and found a nematic phase of the kind discussed
above, which had not been identified previously. Because of the close relation between the nematic order parameter and
the structure factor, we decided to extend the method of reference \cite{Cirillo99} to compute the structure factor
of the model in an external field in the whole phase diagram, i.e. also in the relevant ordered phases. 

Results on the $J_1$-$J_2$ Ising model may be relevant to understand part of the phenomenology of high temperature superconductors,
specially the iron pnictides. For these compounds, a much studied model is the quantum Heisenberg $J_1$-$J_2$ \cite{Chandra1990,Fang2008,
Fernandes2012}. This model was shown to have a Ising-nematic phase driven by spin fluctuations, which break the Z$_4$ symmetry of
the square lattice, without the development of anti-ferromagnetic order \cite{Chandra1990}. Strong spin fluctuations in this
2D system induce a biquadratic or quadrupolar interaction leading to Ising-like behavior in spin space and eventually to the
presence of an Ising-nematic phase. Nevertheless, it is not clear if the quadrupolar coupling is strong enough to apply to the
experimental compounds which show $J_1$-$J_2$ behavior. Another route to nematic order in the pnictides seems to be related with
doping. Recent results of Monte Carlo simulations on a model with magnetic, electronic and orbital degrees of freedom imply that
the nematic phase is enhanced through Fe substitution by impurities, i.e. by introducing quenched disorder and magnetic dilution
in the parent compound \cite{Liang2015}. 
Very recently, Kitada {\em et al.} \cite{Kitada2016} reported on an extensive series of experiments on the layered perovskite
RbLaNb$_2$O$_7$ transformed by substitution of the Rb on the oxyhalides (MCl)LaNb$_2$O$_7$ (M=Mn, Cr, Co), which are two dimensional
antiferromagnets. Neutron
diffraction measurements show the presence of magnetic modulations with wave vectors (0,$\pi$) for the samples with Mn and Co. The
samples with Cr showed instead a ($\pi,\pi$) N\'eel antiferromagnetic structure. Interestingly, 
hysteresis measurements on the (CoCl)LaNb$_2$O$_7$ compound indicate a way to saturation in two steps, as the field is raised. This
is interpreted as Ising-like behavior, in which the striped ground stated is destabilized by a ferromagnetic component by first
flipping half of the antiferromagnetic stripes and at a higher field value the other half is flipped,
leading to the completely saturated state. If confirmed, this is the first compound to show a phenomenology typical of the Ising
$J_1-J_2$ model studied in the present work. 

In the following,
we make a brief discussion of known results on the $J_1$-$J_2$ Ising model, the CVM approach, and compute the structure factor
of the model in presence of an external field in the whole parameter range. We interpret the
results in connection with the recently published phase diagram~\cite{Guerrero15}.
We also identify the disorder variety 
of the model in an external field and discuss the possible exactness of the four point CVM approximation on this variety.
\section{$J_1$-$J_2$ Ising model and the CVM approximation}
\label{J1J2}

The $J_1$-$J_2$ Ising model on the square lattice is defined by the Hamiltonian:
\beq
{\cal H} =J_1\sum_{\left<xy\right>} S_x S_y+J_2\sum_{\left<\left<xy\right>\right>}S_xS_y
- h\,\sum_x  S_x,
\label{HJ1J2}
\eeq
where $\{S_x=\pm 1, x=1\ldots N\}$ are $N$ Ising spin variables and $h$ is an external field.
$\left<xy\right>$ denotes a sum over pairs of nearest-neighbors and  $\left<\left<xy\right>\right>$ a sum over pairs of 
next-nearest-neighbors. In this work we consider $J_1<0$ and $J_2>0$ representing ferromagnetic NN and 
anti-ferromagnetic NNN interactions respectively. The competition ratio is defined by $\kappa=\frac{J_2}{|J_1|}\geq 0$.

\begin{figure}[ht!]
\centering
\includegraphics[scale=1.]{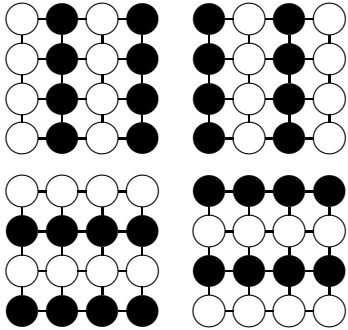}
\caption{Sketch of the striped ground state configurations of the $J_1$-$J_2$ model in the square lattice.}
\label{fig.gs}
\end{figure}

At zero external field the ground state of the model is ferromagnetic
for $\kappa<1/2$ and striped (superantiferromagnetic) if $\kappa>1/2$. The stripe phase has
fourfold degeneracy  as shown in Figure \ref{fig.gs}.
For $h=0$ the model has been extensively studied \cite{Moran93,Moran94,Cirillo99,dosAnjos08,
Kalz09,Kalz11,Jin12,Jin13,Saguia13}. The nature of the thermal phase transition from
the stripes to a disordered phase for $\kappa > 1/2$ was controversial. In the most recent studies combining Monte Carlo
simulations and a series of analytic techniques it has  been established that the line of phase transitions in the 
temperature versus $\kappa$ plane is first order for $1/2 < \kappa < 0.67$ and is continuous with Ashkin-Teller critical
behavior for $\kappa > 0.67$. The critical exponents change continuously in this regime between the 4-state Potts model
behavior at $\kappa=0.67$ to standard Ising criticality for $\kappa \to \infty$~ \cite{Jin12,Jin13}.

For $\kappa>\frac{1}{2}$ and small magnetic fields the ground state is striped. 
Stripe order is eventually destroyed at a critical field value $h_c=\pm 2(J_1+2J_2)$.
 For $h > h_c$ all spins are aligned with the field and the ground state corresponds to a saturated paramagnet.
In a recent work it was found a new equilibrium phase at intermediate fields in the $h$ vs $T$ plane, 
a Ising-nematic phase with uniform magnetization but different nearest-neighbor correlations along the two directions 
of the square lattice, breaking the fourfold rotational symmetry~\cite{Guerrero15}.
At lower fields a second transition takes place to a full stripe phase with broken rotational as well as translational symmetries. 
In order to admit a nematic phase the system must have enhanced fluctuations, this is the reason behind the Ising-nematic phase
originally reported in the Heisenberg $J_1$-$J_2$ model \cite{Chandra1990}. In the 2D Heisenberg model strong fluctuations are
due to the continuous rotational symmetry in spin space. In the Ising model this mechanism is absent, but enhanced fluctuations
can be induced by the switching of a magnetic field. In fact, in a restricted temperature window which induces temperature
fluctuations, the net effect of an external field on the stripe ground state is to favor one of the Ising directions and to weaken
the other. For a suitable intensity the field will be responsible for inducing defects on the stripe pattern and eventually an
instability leading to the loss of anti-ferromagnetic positional order, but still preserving a preferred orientation for the remaining
stripe pattern. This is precisely the signature of the Ising-nematic phase, a phase with broken positional order (or translation
symmetry) but still having orientational order of the stripe pattern on the lattice.

The Cluster Variation Method consists in improving a mean field approximation in a systematic way by summing all the degrees of
freedom within clusters of size $n$ in an exact way. It amounts to extremize the variational free energy:
\beq
F_t=Tr\,(\rho_t H)+k_BT\ Tr\,(\rho_t \ln{\rho_t}),
\eeq
where $Tr$ means a trace or a sum over all the relevant degrees of freedom of the Hamiltonian $H$, and
$\rho_t$ is a trial density matrix which satisfies the normalization constraint $Tr\,\rho_t=1$. 
Details on the method can be found in the large literature on the subject (see e.g. references \cite{Kikuchi51,Sanchez84,An88,Tanaka2002}).

In the case of a system with Ising spins $\{S_i=\pm 1\}$, the reduced density matrix for a cluster of size $n$, $\rho_t^{(n)}$, can be written as
\cite{Cirillo99}:
\beq
\rho_t^{(n)}=2^{-n}\left[1+\sum_{k}\sigma_k\zeta_k \right]
\eeq
where the sum runs over all sub-clusters with $k$ sites within cluster $n$ , $\sigma_k=\prod_{i\in k}S_i$ and the k-point correlation 
functions are defined by $\zeta_k=Tr\, \sigma_k \rho_t^{(k)}$. In this way, instead of optimizing with respect to the reduced densities, the
variational parameters are the k-point correlations $\zeta_k$ which must satisfy:
\beq
\frac{\partial F_t}{\partial \zeta_k}=0.
\label{state.eq}
\eeq
A hierarchy of approximations to the free energy can be constructed in this way. The simplest one corresponds to the 1-point
approximation for the density matrices, the usual mean field approximation. The 2-point approximation is usually called
Bethe-Peierls approximation~\cite{Bethe1935,Tanaka2002}. As discussed in 
\cite{Guerrero15} the minimal cluster able
to capture the emergence of anisotropic nearest-neighbor correlations or spontaneous rotational symmetry
breaking is the four-point or square approximation.

Define the correlation functions:
\beqa
m_x&\equiv & \langle S_x \rangle =Tr\left( S_x\rho_x\right) \nonumber\\
l_{xy}&\equiv& \langle S_xS_y \rangle=Tr \left(S_x S_y \rho_{\left<xy\right>} \right)\nonumber\\
c_{xz}&\equiv&\langle S_xS_z \rangle = Tr \left(S_x S_z \rho_{\left<\left<xz\right>\right>}\right) \nonumber\\
k_{yxw}&\equiv& \langle S_yS_xS_w \rangle =Tr \left(S_y S_x S_w\rho_{\left[yxw\right]} \right)\nonumber\\
d_{xyzw}&\equiv& \langle S_xS_yS_zS_w \rangle =Tr \left(S_x S_y S_z S_w \rho_{^x_y\Box^w_z}\right),\nonumber\\
\eeqa
where the sums over $x$, $\left<xy\right>$, $\left<\left<xy\right>\right>$, $[xyz]$, 
${^x_y \Box^w_z}$ refer to all sites, NN pairs, NNN pairs, 
clusters of three sites and squares respectively.
They are related to the reduced density matrices by:
\beqa
\rho_x&=&\frac{1}{2}\left( 1+m_xS_x\right) \nonumber\\
\rho_{\left<xy\right>}&=&\frac{1}{4}\left(1+m_xS_x+m_yS_y+l_{xy}S_xS_y\right)\nonumber\\
\rho_{\left<\left<xz\right>\right>}&=&\frac{1}{4}\left(1+m_xS_x+m_zS_z+c_{xz}S_xS_z\right)\nonumber\\
\rho_{^x_y\Box^w_z}&=&\frac{1}{16}(1+m_xS_x+m_yS_y+m_zS_z+m_wS_w\nonumber\\
&+&l_{xw}S_xS_w+l_{wz}S_wS_z+l_{zy}S_zS_y+l_{xy}S_xS_y\nonumber\\
&+&c_{xz}S_xS_z+c_{yw}S_yS_w+k_{yxw}S_yS_xS_w \nonumber\\
&+&k_{xwz}S_xS_wS_z+k_{wzy}S_wS_zS_y +k_{zyx}S_zS_yS_x\nonumber\\
&+&d_{xyzw}S_xS_yS_zS_w).
\eeqa
Then, the free energy of the $J_1$-$J_2$ model in the CVM square approximation can be written as~\cite{Cirillo99,Guerrero15}:
\beqa
 F&=&J_1\sum_{\left<xy\right>} l_{xy} 
+J_2\sum_{\left<\left<xy\right>\right>} c_{xy} - h\,\sum_x m_x \nonumber\\
&+&k_B T \left[ \sum_x Tr \left(\rho_x log \rho_x\right)  
 -\sum_{\left<xy\right>}Tr \left(\rho_{\left<xy\right>} log \rho_{\left<xy\right>}\right) \right. \nonumber\\
&+&\left. \sum_{^x_y\Box^w_z} Tr \left(\rho_{^x_y\Box^w_z} log\rho_{^x_y\Box^w_z}\right) \right].
\label{FJ1J2}
\eeqa

After computing the traces one is left with an expression for the variational free energy in terms of a set
of correlation functions representative of the approximation considered. The form of equation (\ref{FJ1J2})
makes clear that up to the pair approximation the two directions in the square lattice enter in a completely
symmetric way, the different pairs of sites are decoupled. It is in the last term of (\ref{FJ1J2}), when square plaquettes are
considered, that the coupling between different directions in space can lead to novel behavior. 

\section{The structure factor and the nematic order parameter}
\label{sf}
In order to detect orientational order it is natural to define orientational or
nematic order parameters. The nematic order parameter was introduced originally
in the study of ordered phases of liquid crystals~\cite{deGPr1998,ChLu1995}. 
More recently, nematic-like order was found to be useful to characterize
orientation of interfaces or space modulations of some physical density, like
electron density in the so-called "electronic liquid-crystals"~\cite{KiFrEm1998} or spin density
in magnetic systems~\cite{AbKaPoSa1995,NiSt2007}. In these systems a nematic order 
parameter can be defined, in analogy with the nematic order parameter of liquid crystals,
as a second-rank symmetric traceless tensor which encodes the $180^o$ rotational symmetry of nematic phases. 
In two dimensions the nematic tensor has only one independent entry which can be written as~\cite{StBa2010}:
\beq
Q= \int d^2k\, k^2\,\cos{(2\theta)}S(\vec k),
\eeq
where $\vec k=(k_x,k_y)$ and $\theta$ is the angle between the local wave vector $\vec k$ and a fixed direction in the plane. It is clear that 
the nematic order parameter amounts to compute a weighted average of the structure factor of the system. The weighting factor $\cos{(2\theta)}$ 
has exactly the symmetry of the nematic phase and then $Q$ will be zero if $S(\vec k)$ is isotropic. Then, {\em the nematic order parameter 
amounts to compute the degree of anisotropy of the structure factor of the system}. Also, because the structure factor is a quantity of primary 
experimental relevance, it is interesting to be able to compute it in the context of the Cluster Variation Method. This has been done in 
a few previous works~\cite{Sanchez82,FineldeFontaine86,Cirillo99}. In reference \cite{Cirillo99} the authors introduced a general 
method for the computation of the structure factor at any level of approximation in the CVM and computed it for the paramagnetic phase of 
the $J_1$-$J_2$ model at zero external field in the four point approximation in the square and simple cubic lattices. 
The starting point is the computation of the two-point connected correlation function:
\beq
C_c(\vec r)\equiv \langle S_0S_{\vec r}\rangle -
\langle S_0\rangle \langle S_{\vec r}\rangle= k_B T \frac{\partial^2F}{\partial h_0 \partial h_{\vec r}} .
\label{Cc}
\eeq
Then, in the square lattice, the structure factor is simply the discrete Fourier transform of $C_c(\vec r)$:
\beq
S(\vec k)=\sum_{r_1,r_2=0}^{L-1} e^{-2\pi i(\vec k\cdot \vec r)/L} C_c(\vec r).
\eeq
In this work, we have extended the method introduced in reference \cite{Cirillo99} in order to compute $S(\vec k)$ also in the ordered phases of the 
$J_1$-$J_2$ model and with the inclusion of a uniform external field $h$. Details of the calculations are presented in the Appendix. 
We are interested in particular in identifying signatures of the nematic phase in the structure factor. The computation in this case, although straightforward, 
is considerably more cumbersome than in the zero field case for the paramagnetic phase, because of the presence of a non zero magnetization 
and the need to consider the symmetry of the ordered phases.  With the aim of searching for purely orientational nematic-like phases, i.e. phases without translational order, 
we minimized the CVM free energy of Eq. (\ref{FJ1J2}) with the following symmetry in the parameters (related to the four points of the elementary square ${^x_y\Box^w_z}$): 
$m_x=m_y$, $m_w=m_z$, $l_{xw}=l_{yz}$, $l_{xy}$, $l_{wz}$, $c$, $k_{yxw}=k_{zyx}$, $k_{xwz}=k_{wzy}$ and $d$. This choice implies possible orientational order along
the $xy$ or vertical direction. Note that local magnetizations on horizontal NN sites are allowed to be
different in sign and also in absolute value. Correspondingly, the NN correlation functions $l_{rs}$ may
be different not only between the horizontal and vertical directions but also between the two vertical ones. 
With these choices the values of NNN correlations $c$ and square correlations $d$ are unique. 
The values of $m_{r}$, $l_{rs}$, $c_{rs}$, $k_{rst}$ and $d$ with $r,s,t=x,y,z,w$ that minimize the variational free energy (\ref{FJ1J2}) were 
calculated numerically for different temperatures ($T$) and external fields ($h$) as described in~\cite{Guerrero15}.

After the symmetry considerations, the structure factor is found to take the form (see the Appendix for details):
\begin{eqnarray}
S(\vec{k})^{-1}&=&2(\gamma_{xx}+\gamma_{ww}) + 8 \gamma_{xw} \cos{k_1} + 4 ( \gamma_{xy}+\gamma_{zw}) \cos{k_2} \nonumber\\
&+& 8 \gamma_{xz} \left[ \cos{(k_1+k_2)} + \cos{(k_1-k_2)}\right]
\label{cinvk}
\end{eqnarray}
where $\vec k=(k_1,k_2)$ and the coefficients $\gamma_{rs}$ are the inverse pair connected correlations as defined in the Appendix. 
They depend on $m_r$, $l_{rs}$, $c_{rs}$. $k_{rst}$ and $d$ with $r,s,t=x,y,z,w$ as in equation (\ref{gammaij}).


\section{Results}
\label{sec:res}
\subsection{Generalized susceptibilities and phase transitions}
In this section we describe the evolution of the structure factor with decreasing magnetic field for $\kappa=0.6$ at a fixed reduced temperature $T/|J_1|=0.67$. 
 Along this line the system goes through two successive 
phase transitions as the external field is lowered:  one from the isotropic to the Ising-nematic phase and another one, at lower field value, 
from the nematic to the stripe phase. The structure factor shows a series of extremes in the $(k_1,k_2)$ plane, with heights which evolve 
with the temperature and magnetic field. These extremes are located at wave-vectors $(0,0)$, $(0,\pm \pi)$ and $(\pm \pi,0)$. 
The heights of the extremes correspond to generalized susceptibilities:
\beq
\chi(k_1,k_2) = \beta S(k_1,k_2).
\eeq

\begin{figure}[ht!]
\centering
\includegraphics[scale=0.4]{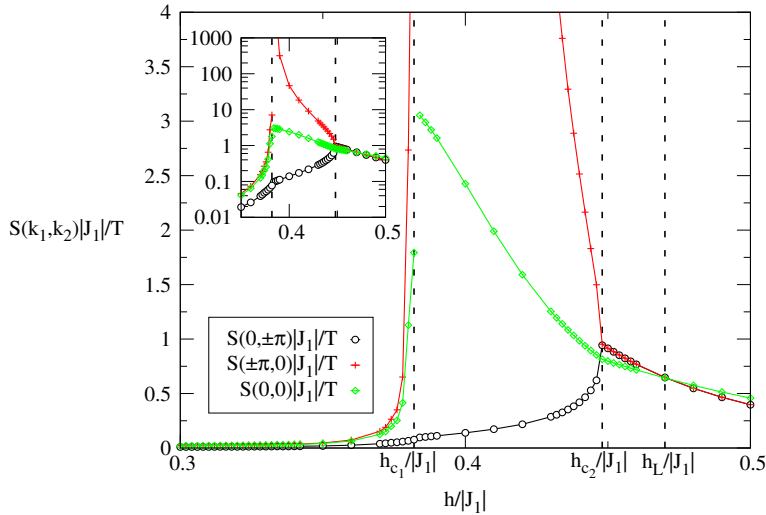}
\caption{(Color online) Generalized susceptibilities $\chi(k_1,k_2)$ at $\vec k=(0,0),(0\pm \pi)$ and $(\pm \pi,0)$ for $T/|J_1|=0.67$ as function of
external field. Inset: zoom around the highest peak in logarithmic scale.}
\label{fig:suscep}
\end{figure}

Typically, at a second order phase transition the generalized susceptibility displays critical behavior at particular values of the wave vector, 
diverging at the critical point. At a first order phase transitions it suffers a discontinuous jump. The susceptibilities corresponding to the 
five extremes are shown in Figure \ref{fig:suscep}.
%
%
For large fields $h /|J_1| > 0.45$ the system is in a homogeneous paramagnetic state. In this region $\gamma_{xx}=\gamma_{ww}\equiv \gamma$,
$\gamma_{xw}=\gamma_{xy}=\gamma_{zw}\equiv \gamma_1$ and $\gamma_{xz}\equiv \gamma_2$ and the structure factor has the same form as in the zero
field case~\cite{Cirillo99}.
For $h /|J_1| > 0.47$ the maximum corresponds to the peak at the origin $\vec k=(0,0)$ (green line). The other extremes (saddles,
 red and black lines) are located on the axes at the border of the first Brioullin zone, $\vec k=(\pm \pi,0)$ and $\vec k=(0,\pm \pi)$,
 have equal heights and  cross  the green line at $h_L/|J_1|=0.47$.  This point is a ``Lifshitz point'', where the solutions at non zero
wave-vector become metastable. At this point the inverse pair correlations satisfy $\gamma_1=-2\gamma_2$.

\begin{figure}[ht!]     
    \begin{subfigure}[b]{0.5\columnwidth}
      \includegraphics[width=\columnwidth]{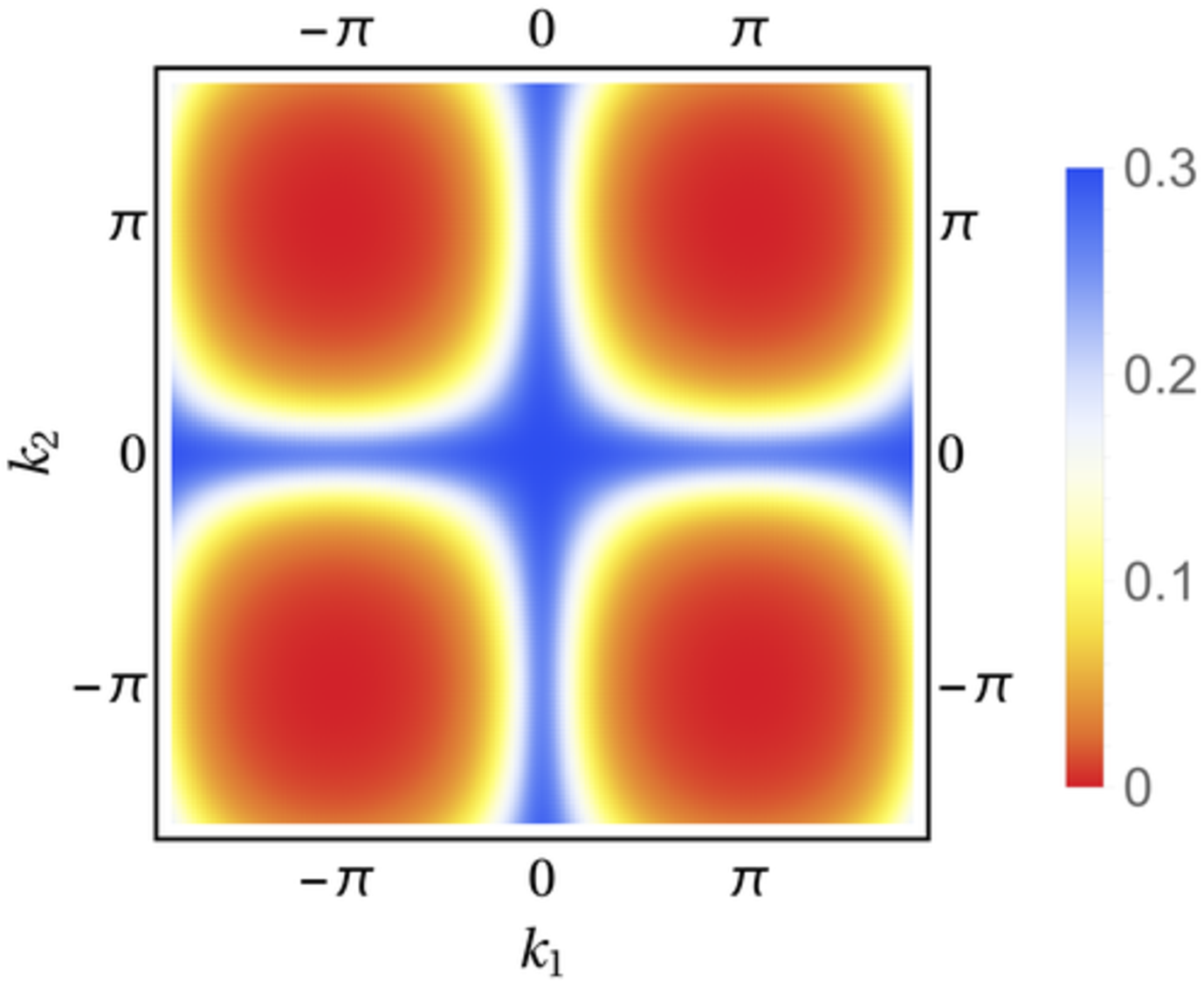}
      \caption{$h/|J_1|=0.5$ (paramagnetic phase). Note the dominant maximum at the origin.}
      \label{fig:dp_h05}
    \end{subfigure}
    \begin{subfigure}[b]{0.5\columnwidth}
      \includegraphics[width=\columnwidth]{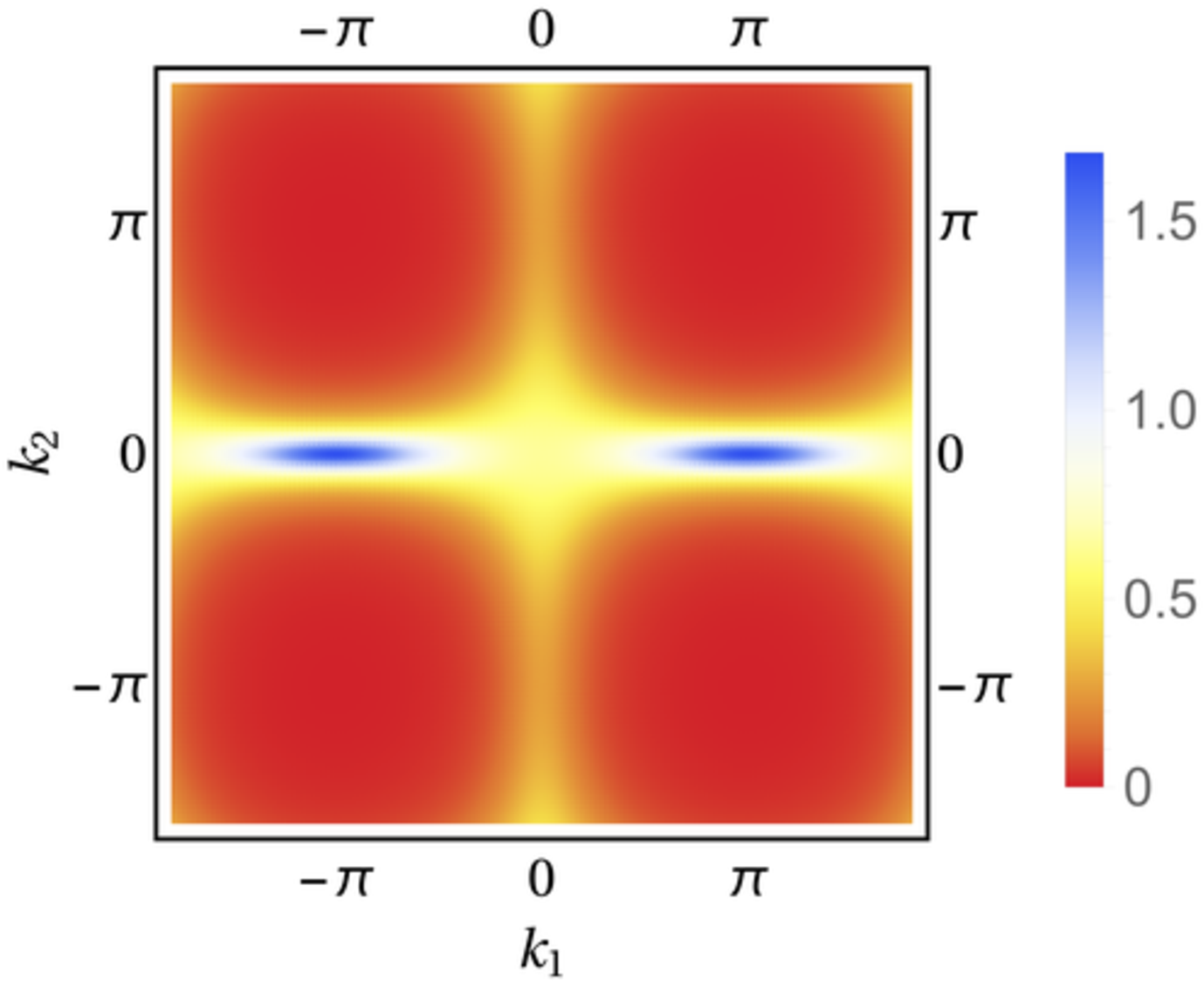}      		  
    \caption{$h/|J_1|=0.44$ (nematic phase). Note the dominant maximum at $\vec k=(\pm \pi,0)$.}
      \label{fig:dp_h044}
    \end{subfigure}         
\begin{center}
    \begin{subfigure}[b]{0.5\columnwidth}
      \includegraphics[width=\columnwidth]{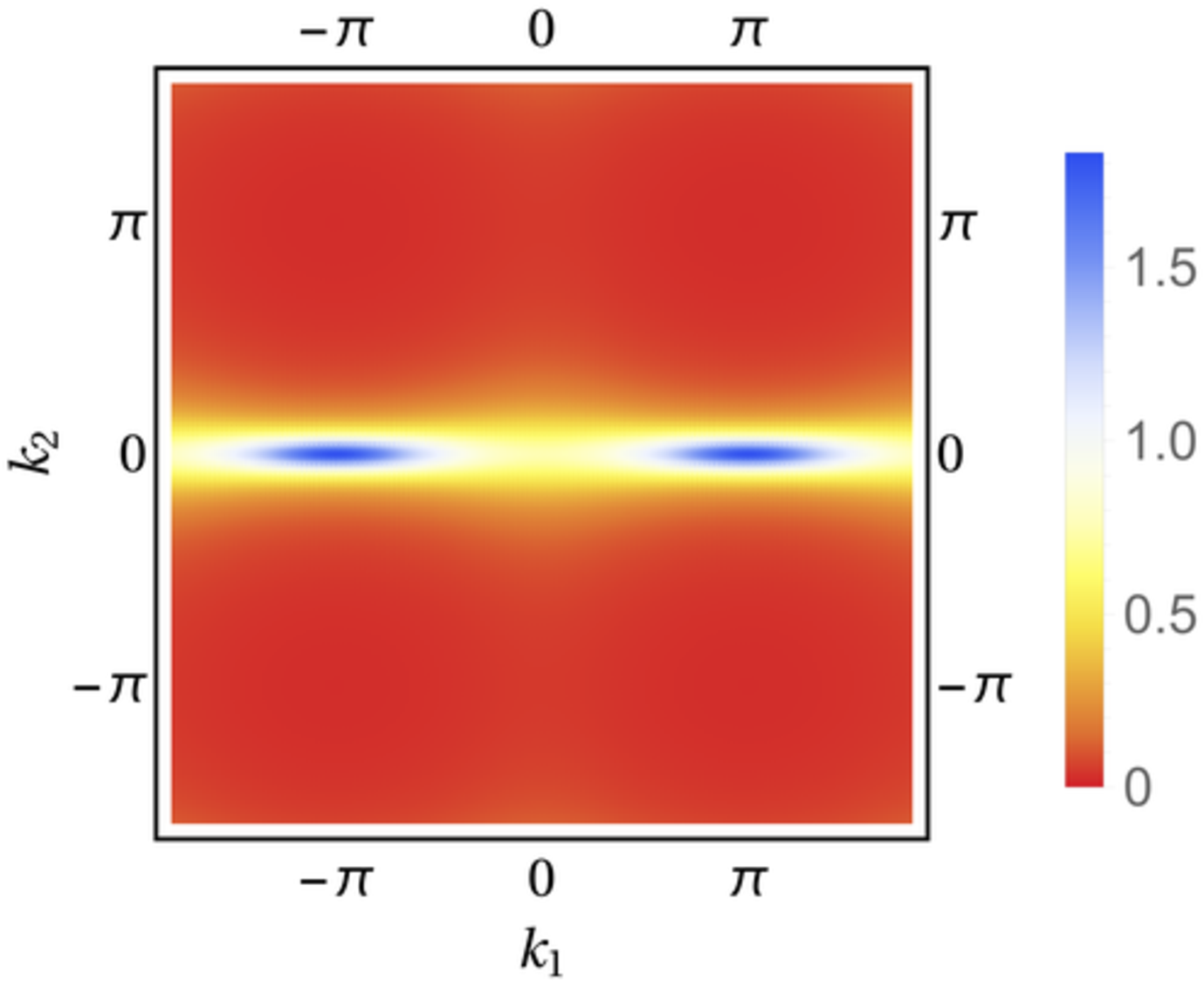}
      \caption{$h/|J_1|=0.38$ (deep in the stripe phase).}
	  \label{fig:dp_h038}
   \end{subfigure}
    \end{center}
\vspace{-2mm}    
 \caption{(Color online) Density plot of the structure factor for $T/|J_1|=0.67$ and different reduced magnetic fields.}
 \label{SvskK=0.6h}
    \end{figure}

A density plot of the structure factor for $h/|J_1|=0.5$ is shown in Fig. \ref{fig:dp_h05}.
The second remarkable fact is that the absolute maxima for $h<h_L$ for $\vec k=(\pm \pi,0)$  and $\vec k=(0,\pm \pi)$ have equal 
height until $h_{c2}/|J_1|=0.45$ where they bifurcate. 
In the whole sector $h/ |J_1| > 0.45$ 
the spatial distribution of the magnetization has $Z_4$ symmetry, characteristic of the square lattice. At $h_{c2}/|J_1|=0.45$ there 
is a {\em spontaneous breaking of $Z_4$ symmetry}, to a phase with a lower, $Z_2$ symmetry. This phase transition, which was shown 
to be continuous in \cite{Guerrero15}, is a paramagnetic to Ising nematic phase transition. Note that the generalized 
susceptibilities represent fluctuations of the magnetization and not of the nematic order parameter and then, although the transition 
is second order, it is not accompanied by a divergence of the magnetic susceptibility. In order to study the divergence of the nematic 
susceptibility it would be necessary to go beyond two point correlations and consider their own fluctuations, i.e. four-point correlations. 
A density plot of the structure factor in the Ising-nematic phase for $h/|J_1|=0.44$ is shown in Fig. \ref{fig:dp_h044}.

As can be seen in Fig. \ref{fig:suscep} the secondary peaks at $\vec k=(0,\pm \pi)$ decrease rapidly for $h<h_{c2}$, while the peak at the 
origin grows steadily although at a much lower rate than the peaks at $(\pm \pi,0)$. Upon further lowering the field
a second singularity appears at a field value $h_{c1}/|J_1|=0.382$ (see the inset in Fig. \ref{fig:suscep}). At this point the susceptibility at $(\pm \pi,0)$ 
changes discontinuously while the peak at $(0,\pm \pi)$ is negligibly small. This transition is not accompanied by a symmetry 
breaking, instead it is a first order transition in which the stripe (translational) order parameter jumps from zero to a finite value for 
$h<h_{c1}$, as originally found and discussed in \cite{Guerrero15}. 
In Fig. \ref{fig:dp_h038} we show a density plot of the structure factor just below the critical field $h/|J_1|=0.38\leq h_{c1}/|J_1|$. 
Our results correspond to the expectations for an Ising-nematic phase, see e.g. figure 2 in \cite{Fernandes2012} and figures 7 and 8
in \cite{Liang2015}.

\subsection{Disorder line}
\begin{figure}[ht!]
\centering
\includegraphics[scale=0.4]{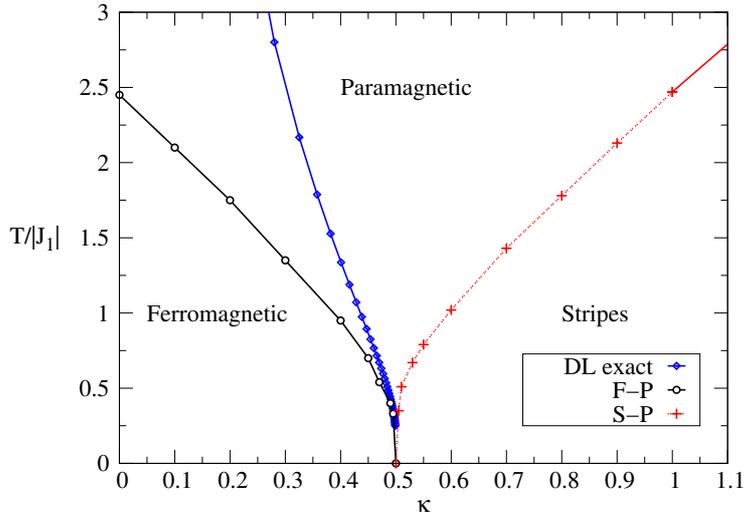}
\caption{(Color online) Phase diagram for $h=0$ from the four point approximation of the CVM. The exact disorder line is also shown.}
\label{fig:disline}
\end{figure}

It is well known that many one and two dimensional frustrated systems have a disorder variety in parameter space which crosses the disordered phase. 
At one side of the disorder variety 
the pair correlations show a monotonic exponential decay with distance, while on the other side the correlations show a modulation or a sinusoidal
decay, typical of the presence of frustrated or competing interactions~\cite{Stevenson1970-1,Stevenson1970-2,Enting1977,Rujan1984}. 
On the disorder variety the pair correlation function usually
factorizes along two perpendicular directions $C(r_x,r_y)=g(r_x)g(r_y)$, showing typical one dimensional behavior which in turn allows to obtain,  in many cases, the
exact solution of the model. This is the case of the $J_1$-$J_2$ Ising model on the square lattice. For the zero field case, it was shown that
the four-point CVM approximation yields the exact solution of the model on the disorder variety~\cite{Pelizzola2000}. In particular, 
$g(u)=\exp{\{-|u|/\xi\}}$ with $\xi=-1/\ln{(l)}$, $l$ being the nearest-neighbor correlation. 
This implies that the structure factor has a similar one dimensional factorization $S(k_x,k_y)=S_1(k_x)S_1(k_y)$, where
$S_1(k_i)=\sinh{(1/\xi)}/[\cosh{(1/\xi)}-\cos{k_i}]$. Given this simple behavior, it is easy to show that on the disorder variety the inverse pair
correlations satisfy the constraint $\gamma_1^2=-\gamma \gamma_2$. This is an exact relation for the zero field case which, when written in terms
of $T/|J_1|$ and $J_2/|J_1|$, leads to the disorder line shown in Figure \ref{fig:disline}.
When an external field is present the general form of the structure factor is given by (\ref{cinvk}). Nevertheless, one expects that the disorder line
should continue to exist in the disordered region with $\kappa < 0.5$, where there is a unique paramagnetic phase with finite magnetization. In this context, it
is natural to expect that the structure factor should have the same simple form of the $h=0$ case, with only three relevant parameters
$\gamma,\ \gamma_1$ and $\gamma_2$. Then, assuming a one dimensional factorization on the disorder variety, as in the $h=0$ case, the same relation
$\gamma_1^2=-\gamma \gamma_2$ defines a disorder surface, where in this case each $\gamma$ is a function of $T/|J_1|,\  J_2/|J_1|$ and $h/|J_1|$.
The similar form of the structure factor in the disordered phases of the zero field and finite field cases suggests that also for finite $h$ the
relation which defines the disorder variety should be exact. The existence of an exact solution for a system which displays a nematic-like phase,
 even limited to the disordered region of the phase diagram, is an interesting possibility.
This is an open question which deserves further study and is beyond the scope of the present work.

\section{Conclusions}
\label{conc}
We have computed the pair correlations and structure factor of the
$J_1$-$J_2$ square lattice Ising model in an external field within the Cluster Variation Method. 
Our motivation was the analysis of the recently reported Ising-nematic phase in a sector of
the $T-h$ phase diagram of the model. Because the nematic order parameter is a function of the structure factor then the latter should show 
clear signatures of the presence of a nematic-like phase. Considering rather general symmetry conditions for the values of the local
magnetizations in an elementary square of the lattice, we applied the four point approximation in the Cluster Variation Method, which is
the minimal approximation capable of detecting the presence of an orientational phase of nematic character. We showed that the results
for the structure factor are in agreement with the corresponding phase diagram reported in  \cite{Guerrero15} and that, although the nematic susceptibility is not
directly related to the structure factor, the presence of the Ising-nematic phase is clearly evidenced in its behavior.

It was shown that the disorder variety of the model is defined by a constraint between the inverse correlations in an elementary square and
that the form of the constraint is the same for the cases with or without external field. This, together with the exactness of the four point
approximation on the disorder variety at zero field makes it plausible that also for finite field the approximation should be exact on
the disorder variety, a point that deserves further study.

The observation of nematic-like phases in condensed matter systems, like ultrathin ferromagnetic films and electronic liquids,  is 
growing rapidly in recent years and experimental determination of structure factors and Fermi surfaces in fermionic systems is being
increasingly reported in studies of low dimensional magnetic systems at the nanoscale and high temperature superconductor systems, to
cite two important examples. Then, the analytic determination of the structure factor in suitable approximations like the ones accessible
within the Cluster Variation Method is a valuable tool to compare with computer simulation studies and experimental results on these systems.

 \section*{Acknowledgment}
A.G.D. and D.A.S. acknowledge partial financial support by Conselho Nacional de Desenvolvimento
Cient\'{\i}fico e Tecnol\'ogico (CNPq), Brazil fellowship 303917/2013-0.
\appendix
\section{Connected correlations within the four-point approximation}
The general method to compute pair correlation functions in the CVM has been studied by some authors \cite{Sanchez82,FineldeFontaine86,Ricci12,Cirillo99}. 
The method consists in introducing an external magnetic field in the variational potential and expressing the correlation functions in terms of successive derivatives 
of the free energy with respect to the magnetic field, as in equation (\ref{Cc}). However, it is easier to compute the inverse connected correlation function:
\begin{equation}
\beta \left( \frac{\partial h_i}{\partial m_j}\right).
\end{equation}
In the four-site approximation, the expression for the inverse correlation function can be obtained from the state equation for the magnetization, 
differentiating the variational free energy (\ref{FJ1J2}) with respect to $m_i$:
\begin{eqnarray}
\beta h_i&=&-\frac{1}{4}\sum_{\left<iy\right>}Tr( S_i log\rho_{\left<iy\right>})+
\frac{1}{16}\sum_{^i_y\Box^w_z}Tr(S_i log \rho_{^i_y\Box^w_z}) \nonumber\\
&+& \frac{1}{2}Tr(S_i log \rho_i),
\label{hi}
\end{eqnarray}

and then differentiating with respect to $m_j$:

\begin{eqnarray}
&&\beta \, \frac{\partial h_i}{\partial m_j}=\frac{1}{2}\,{\rm tr}\, \frac{S_i}{\rho_i}\,\frac{\partial \rho_i}{\partial m_j}-\frac{1}{4}\,\sum_{\left<ij\right>}\,{\rm tr}\, \frac{S_i}{\rho_{ij}} \,\left[\frac{\partial \rho_{ij}}{\partial m_j} + \frac{\partial \rho_{ij}}{\partial l_{ij}} \, \frac{\partial l_{ij}}{\partial m_j} \right] 
+ \frac{1}{16} \, \sum_{^i_j\Box^w_z} \,{\rm tr}\,\frac{S_i}{\rho_{^i_j\Box^w_z}}\left[\frac{\partial \rho_{^i_j\Box^w_z}}{\partial m_j} + \frac{\partial \rho_{^i_j\Box^w_z}}{\partial l_{ij}}\;\frac{\partial l_{ij}}{\partial m_j} \right.  \nonumber\\
&&+ \left. \frac{\partial \rho_{^i_j\Box^w_z}}{\partial l_{iw}}\;\frac{\partial l_{iw}}{\partial m_j} 
+ \frac{\partial \rho_{^i_j\Box^w_z}}{\partial l_{zw}}\;\frac{\partial l_{zw}}{\partial m_j}  + 
 \frac{\partial \rho_{^i_j\Box^w_z}}{\partial l_{jz}}\;\frac{\partial l_{jz}}{\partial m_j} 
+ \frac{\partial \rho_{^i_j\Box^w_z}}{\partial c_{iz}}\;\frac{\partial c_{iz}}{\partial m_j} + \frac{\partial \rho_{^i_j\Box^w_z}}{\partial c_{jw}}\;\frac{\partial c_{jw}}{\partial m_j} 
+  \frac{\partial \rho_{^i_j\Box^w_z}}{\partial k_{wij}}\;\frac{\partial k_{wij}}{\partial m_j} \right.  \nonumber\\
&&+ \left. \frac{\partial \rho_{^i_j\Box^w_z}}{\partial k_{zwi}}\;\frac{\partial k_{zwi}}{\partial m_j} + 
 \frac{\partial \rho_{^i_j\Box^w_z}}{\partial k_{ijz}}\;\frac{\partial k_{ijz}}{\partial m_j} + \frac{\partial \rho_{^i_j\Box^w_z}}{\partial k_{jzw}}\;\frac{\partial k_{jzw}}{\partial m_j} +  \frac{\partial \rho_{^i_j\Box^w_z}}{\partial d_{ijzw}}\;\frac{\partial d_{ijzw}}{\partial m_j} \right].
\label{himj}
\end{eqnarray}

In the general case, all the derivatives $\frac{\partial l_{xy}}{\partial m_x}$, $\frac{\partial c_{xz}}{\partial m_x}$, $\frac{\partial k_{xyz}}{\partial m_x}$ and $\frac{\partial d_{xyzw}}{\partial m_x}$ 
are different of zero. To compute the derivatives  is not an easy task. One way, as is mentioned in reference \cite{Cirillo99}, is by differentiating the variational 
free energy with respect to $l_{xy}$, $c_{xz}$, $k_{xyz}$ and $d_{xyzw}$:

\begin{tabular}{p{3cm}p{4cm}}
{\begin{eqnarray*}
&\frac{\partial \cal F}{\partial l_{xy}}=0 \\    &\frac{\partial \cal F}{\partial c_{xz}}=0
\end{eqnarray*}}
&
{\begin{eqnarray}
&\frac{\partial \cal F}{\partial k_{xyz}}=0 \nonumber\\
&\frac{\partial \cal F}{\partial d_{xyzw}}=0,
\end{eqnarray}}
\end{tabular}

\noindent and then, with respect to $m_j$, $j=x,y,z,w$:

\begin{tabular}{p{3cm}p{4cm}}
{\begin{eqnarray*}
&\frac{\partial}{\partial m_j}\left(\frac{\partial \cal F}{\partial l_{xy}}\right)=0 \\    &\frac{\partial}{\partial m_j}\left(\frac{\partial \cal F}{\partial c_{xz}}\right)=0
\end{eqnarray*}}
&
{\begin{eqnarray}
&\frac{\partial}{\partial m_j}\left(\frac{\partial \cal F}{\partial k_{xyz}}\right)=0 \nonumber\\
&\frac{\partial}{\partial m_j}\left(\frac{\partial \cal F}{\partial d_{xyzw}}\right)=0.  \nonumber\\
\end{eqnarray}}
\end{tabular}

Giving the symmetry considerations in the $J_1-J_2$ model: $m_x=m_y$, $m_w=m_z$, $l_{xw}=l_{yz}$, $l_{xy}$, $l_{wz}$, $c$, $k_{yxw}=k_{zyx}$, $k_{xwz}=k_{wzy}$ and $d$, we found a system of twenty two linear equations and twenty two variables  $l_1,...,l_8$, $c_1,...,c_4$, $k_1,...,k_8$ and $d_1,d_2$, where:

\begin{tabular}{p{2cm}p{2cm}p{2cm}p{2cm}}
{\begin{eqnarray*}
&l_1=\frac{\partial l_{xy}}{\partial m_x}\\
&l_5=\frac{\partial l_{xy}}{\partial m_w} \\
&c_1=\frac{\partial c_{xz}}{\partial m_x} 
\end{eqnarray*}
}
&
{\begin{eqnarray*}
&l_2=\frac{\partial l_{xw}}{\partial m_x} \\
&l_6=\frac{\partial l_{xw}}{\partial m_w} \\
&c_2=\frac{\partial c_{yw}}{\partial m_x} 
\end{eqnarray*}}
&
{\begin{eqnarray*}
&l_3=\frac{\partial l_{zw}}{\partial m_x}  \\
&l_7=\frac{\partial l_{zw}}{\partial m_w} \\ 
&c_3=\frac{\partial c_{xz}}{\partial m_w} 
\end{eqnarray*}}
& 
{\begin{eqnarray*}
&l_4=\frac{\partial l_{yz}}{\partial m_x}  \\
&l_8=\frac{\partial l_{yz}}{\partial m_w}  \\
&c_4=\frac{\partial c_{yw}}{\partial m_w} 
\end{eqnarray*}}
\end{tabular}

\begin{tabular}{p{2cm}p{2cm}p{2cm}p{2cm}}
{\begin{eqnarray*}
&k_1=\frac{\partial k_{yzw}}{\partial m_x}\\
&k_5=\frac{\partial k_{yzw}}{\partial m_w} \\
&d_1=\frac{\partial d_{xyzw}}{\partial m_x}
\end{eqnarray*}
}
&
{\begin{eqnarray*}
&k_2=\frac{\partial k_{xyz}}{\partial m_x} \\
&k_6=\frac{\partial k_{xyz}}{\partial m_w}\\
&d_2=\frac{\partial d_{xyzw}}{\partial m_w}
\end{eqnarray*}}
&
{\begin{eqnarray*}
&k_3=\frac{\partial k_{zwx}}{\partial m_x}  \\
&k_7=\frac{\partial k_{zwx}}{\partial m_w}
\end{eqnarray*}}
& 
{\begin{eqnarray}
&k_4=\frac{\partial k_{wxy}}{\partial m_x}  \nonumber\\
&k_8=\frac{\partial k_{wxy}}{\partial m_w} \nonumber\\
\label{lckd}
\end{eqnarray}}
\end{tabular}

Differentiating (\ref{hi}) with respect to $m_j$, with $i,j$ nearest neighbors, we found that derivatives of the kind $\frac{\partial l_{xy}}{\partial m_j}$, 
$\frac{\partial c_{xz}}{\partial m_j}$, $\frac{\partial k_{xyz}}{\partial m_j}$ and $\frac{\partial d_{xyzw}}{\partial m_j}$ where $j$ $\not \in$ ${^x_y\Box^w_z}$
appear. Following the procedure described before, we found a set of twenty two linear equations and twenty two variables  
$l_1',...,l_8'$, $c_1',...,c_4'$, $k_1',...,k_8'$ and $d_1',d_2'$ similar to (\ref{lckd}) with $m_j$ external to the plaquette. All these derivatives are zero.
Collecting all the pieces, from (\ref{himj}), the inverse correlation functions are given by:

\begin{equation}
\beta \, \left(\frac{\partial h_i}{\partial m_j}\right)=
  \begin{cases}
     \gamma_{xx}+\gamma_{yy}+\gamma_{zz}+\gamma_{ww} & i= j  \\
    4 \gamma_{xw} & \left<ij\right>_h \\
    2 \gamma_{xy}+2\gamma_{zw} & \left<ij\right>_v \\
     4 \gamma_{xz} & \left<\left<ij\right>\right> \\
     0 & \text{otherwise}, 
   \end{cases}
\label{InvPairCor}
\end{equation}

where the derivatives $\beta \left(\frac{\partial h_i}{\partial m_j}\right)$ are the inverse self correlations ($i= j$), NN inverse correlations in the horizontal ($\left<ij\right>_h$) 
and vertical ($\left<ij\right>_v$) direction,  and inverse correlation for NNN  ($\left<\left<ij\right>\right> $), respectively. 
All other correlation functions are zero. 

The coefficients $\gamma_{ij}$ are given by: 

\begin{eqnarray}
\gamma_{xx}&=&\beta \, \frac{\partial h_x}{\partial m_x} \nonumber\\ 
&=&\frac{1}{16} \, R_1 -\frac{1}{32}(Q_5+Q_6+Q_1 \, l_1+Q_2 \, l_2)+\frac{1}{16^2}\left(P_{10}+P_1 \, l_1+P_2 \, l_2+P_8 \, l_3 \right. \nonumber\\
&&\left.+P_7 \, l_4+P_2 \, c_1+P_7 \, c_2+P_9 \, k_1 + P_5 \, k_2 +P_4 \, k_3+P_6 \, k_4 + P_8 \, d_1\right) \nonumber\\
\gamma_{ww}&=&\beta \, \frac{\partial h_w}{\partial m_w} \nonumber\\ 
&=&\frac{1}{16}\, R_2 -\frac{1}{32}(Q_6+Q_7+Q_3 \, l_6+Q_4 \, l_7)+\frac{1}{16^2}\left(P_{10}+P_7 \, l_5+P_1 \, l_6+P_2 \, l_7 \right. \nonumber\\
&&\left.+P_8 \, l_8+P_8 \, c_3+P_1 \, c_4+P_5 \, k_5 + P_9 \, k_6 +P_6 \, k_7+P_3 \, k_8 + P_7 \, d_2\right) \nonumber\\
\gamma_{xw}&=&\beta \frac{\partial h_x}{\partial m_w} \nonumber\\ 
&=&-\frac{1}{2}\, \frac{1}{32}(Q_{10}+Q_2 \,l_6)+\frac{1}{2}\,\frac{1}{16^2}\left(P_{5}+P_1 \, l_5+P_2 \, l_6+P_8 \, l_7 
+P_7 \, l_8+P_2 \, c_3 \right. \nonumber\\
&&\left.+P_7 \, c_4 + P_9\,k_5 + P_5 \, k_6 + P_4 \, k_7+P_6 \, k_8 + P_8 \, d_2\right) \nonumber\\
\gamma_{xy}&=&\beta \frac{\partial h_x}{\partial m_y} \nonumber\\ 
&=&-\frac{1}{2}\, \frac{1}{32}(Q_{8}+Q_1 \,l_1)+\frac{1}{2}\,\frac{1}{16^2}\left(P_{3}+P_1 \, l_1+P_7 \, l_2+P_8 \, l_3 
+P_2 \, l_4+P_7 \, c_1 \right. \nonumber\\
&&\left.+P_2 \, c_2 + P_4\,k_1 + P_6 \, k_2 + P_9 \, k_3+P_5 \, k_4 + P_8 \, d_1\right) \nonumber\\
\gamma_{xz}&=&\beta \frac{\partial h_x}{\partial m_z} \nonumber\\ 
&&=\frac{1}{2}\, \frac{1}{16^2}\left(P_{6}+P_1 \, l_5+P_7 \, l_6+P_8 \, l_7 
+P_2 \, l_8+P_2 \, c_4 \right. \nonumber\\
&&\left.+P_7 \, c_3 + P_4\,k_5 + P_6 \, k_6 + P_9 \, k_7+P_5 \, k_8 + P_8 \, d_2\right), 
\label{gammaij}
\end{eqnarray}

\noindent with:

\begin{tabular}{p{2.5cm}p{2.5cm}p{2.5cm}p{2.5cm}}
{\begin{eqnarray*}
P_1 =& {\rm tr} \, \left( \frac{S_x}{\rho_{^x_y\Box^w_z}}  \right)\\
= &{\rm tr} \, \left( \frac{S_y}{\rho_{^x_y\Box^w_z}}\right) \\
P_2 =& {\rm tr} \, \left( \frac{S_w}{\rho_{^x_y\Box^w_z}}\right)\\
=& {\rm tr} \, \left( \frac{S_z}{\rho_{^x_y\Box^w_z}}\right)\\
\end{eqnarray*}}
&
{\begin{eqnarray*}
P_3  = &{\rm tr} \, \left( \frac{S_x S_y}{\rho_{^x_y\Box^w_z}} \right)\\
P_4 = &{\rm tr} \, \left( \frac{S_w S_z}{\rho_{^x_y\Box^w_z}}\right)\\
P_5 =&{\rm tr} \, \left( \frac{S_x S_w}{\rho_{^x_y\Box^w_z}}\right)\\
P_6  =& {\rm tr} \, \left( \frac{S_x S_z}{\rho_{^x_y\Box^w_z}}\right) \\
=& {\rm tr} \, \left( \frac{S_y S_w}{\rho_{^x_y\Box^w_z}}\right)\\
\end{eqnarray*}}
&
{\begin{eqnarray*}
P_7  =& {\rm tr} \, \left( \frac{S_x S_y S_z}{\rho_{^x_y\Box^w_z}}\right)\\
= &{\rm tr} \, \left( \frac{S_w S_x S_y}{\rho_{^x_y\Box^w_z}}\right) \\
P_8  =& {\rm tr} \, \left( \frac{S_y S_z S_w}{\rho_{^x_y\Box^w_z}}\right)  \\ 
=& {\rm tr} \, \left( \frac{S_z S_w S_x}{\rho_{^x_y\Box^w_z}}\right)\\
P_9  =& {\rm tr} \, \left( \frac{S_x S_y S_z S_w}{\rho_{^x_y\Box^w_z}}\right)\\
P_{10}= &{\rm tr} \, \left( \frac{1}{\rho_{^x_y\Box^w_z}}\right)\\
\end{eqnarray*}}
\end{tabular}

\begin{tabular}{p{2.5cm}p{2.5cm}p{2.5cm}}
{\begin{eqnarray*}
Q_1=&{\rm tr} \, \left( \frac{S_x}{\rho_{xy}}\right)\\
=&{\rm tr} \, \left( \frac{S_y}{\rho_{xy}}\right)\\
Q_2=&{\rm tr} \, \left( \frac{S_w}{\rho_{xw}}\right)\\
=&{\rm tr} \, \left( \frac{S_x}{\rho_{xw}}\right)\\
Q_3=&{\rm tr} \, \left( \frac{S_x}{\rho_{xw}}\right)\\
\end{eqnarray*}}
&
{\begin{eqnarray*}
Q_4=&{\rm tr} \, \left( \frac{S_z}{\rho_{zw}}\right)\\
=&{\rm tr} \, \left( \frac{S_w}{\rho_{zw}}\right)\\
Q_5  =& {\rm tr} \, \left( \frac{1}{\rho_{xy}} \right) \\ 
Q_6  =&{\rm tr} \, \left( \frac{1}{\rho_{xw}} \right)\\
 =&{\rm tr} \, \left( \frac{1}{\rho_{yz}} \right)\\
Q_7  =&{\rm tr} \, \left( \frac{1}{\rho_{zw}}\right)\\
Q_8=&{\rm tr} \, \left( \frac{S_x S_y}{\rho_{xy}}\right)\\
\end{eqnarray*}}
&
{\begin{eqnarray*}
Q_9=&{\rm tr} \, \left( \frac{S_z S_w}{\rho_{zw}}\right)\\
Q_{10}=&{\rm tr} \, \left( \frac{S_x S_w}{\rho_{xw}}\right)\\
R_1=&{\rm tr} \, \left( \frac{1}{\rho_{x}}\right)\\
=&{\rm tr} \, \left( \frac{1}{\rho_{y}}\right)\\
R_2=&{\rm tr} \, \left( \frac{1}{\rho_{w}}\right)\\
=&{\rm tr} \, \left( \frac{1}{\rho_{z}}\right).\\
\end{eqnarray*}}
\end{tabular}

Finally, the discrete Fourier transform of the inverse correlation is:

\begin{eqnarray}
S(\vec{k})^{-1}&=&\beta \, \sum_{r_1,r_2=0}^{L-1} e^{-2\pi i(\vec k\cdot \vec r)/L}  \left(\frac{\partial h_i}{\partial m_j}\right)\nonumber\\
&=&2(\gamma_{xx}+\gamma_{ww}) + 8 \gamma_{xw} \cos{k_1} + 4 ( \gamma_{xy}+\gamma_{zw}) \cos{k_2} \nonumber\\ 
&+& 8 ( \gamma_{xz}) \left[ \cos{(k_1+k_2)} + \cos{(k_1-k_2)} \right].
\end{eqnarray}


\end{document}